# The Interestingness Tool for Search in the Web

e-print

**Iaakov Exman, Gilad Amar and Ran Shaltiel**

Software Engineering Department,
The Jerusalem College of Engineering – Azrieli
POB 3566, Jerusalem, 91035, Israel
`iaakov@jce.ac.il, {amargilad,mmmmsho}@gmail.com`

Categories and subject Descriptors

H.3.3 [Information Search and Retrieval]: Retrieval models, Search Process {1998 ACM Classification}

Information Systems → Information Retrieval → Retrieval models and ranking
→ Novelty in information retrieval {CCS 2012}

**Abstract.** Interestingness – as the composition of Relevance and Unexpectedness – has been tested by means of Web search cases studies and led to promising results. But for thorough investigation and routine practical application one needs a flexible and robust tool. This work describes such an Interestingness based search tool, its software architecture and actual implementation. One of its flexibility traits is the choice of Interestingness functions: it may work with Match-Mismatch and Tf-Idf, among other functions. The tool has been experimentally verified by application to various domains of interest. It has been validated by comparison of results with those of commercial search engines and results from differing Interestingness functions.

# 1  Introduction

Commercial web search engines are quite powerful and have achieved widespread usage. Typically one starts interactive search with a set of input keywords, and after a few search cycles, by gradual modifications of the input set, one halts search with a result set considered satisfactory by some subjective criteria – not explicitly formulated.

The introduction of the idea of Interestingness within web search allows one to achieve two different benefits:

- a- *Domain Relevance* – to focus results on a certain domain of interest, disambiguating domain overlaps;
- b- *Unexpectedness* – to offer explicit quantitative criteria for the unexpectedness aspects of interestingness.

This work embedded these potential benefits into a flexible and robust search tool, and validated the tool in various dimensions.

Next one finds references to related work.

## 1.1  Related Work

The literature on measures of Interestingness is very extensive and appeared in a variety of contexts. Here we provide selected pointers to related work.

An introductory survey on Interestingness measures can be found in references McGarry [10] and Klosgen and Zytkow [8].

More specific papers on Interestingness include Piatetsky-Shapiro and Matheus, [12] and Tuzhilin [13]. The latter refers to integration of different measures of Interestingness. Unexpectedness is explicitly mentioned as a measure of interestingness for knowledge discovery by Padmanabhan and Tuzhilin [11]. Interestingness within Web search is discussed e.g. in Lin, Etzioni and Fogarty [9].

Data mining is a field in which Interestingness plays an important role. See e.g. the papers by Geng, L. and Hamilton [3], [4] and the book by Guillet and Hamilton [5]. Another potential source of Interestingness measures is the field of Exploratory Search. As an example, see the paper by Webb [14].

There have been proposals of interestingness related tools – see e.g. Huynh et al. [6], [7].

Potential applications of interestingness include, among others, Web search of potentially novel medical drug leads (e.g. [2]).

In the remaining of the paper we shortly review Interestingness measures (section 2), describe the tool's software architecture (section 3), deal with the tool's implementation (section 4), describe in detail experiments done to validate the tool (section 5) and conclude with a discussion (section 6).

## 2  Interestingness Measures

Interestingness has been defined by Exman [1] in terms of composition of two functions, relevance *R* and unexpectedness *U* as in the next formula:

$$Int = R \circ U \qquad (1)$$

where *Int* is the Interestingness and the operator ° stands for composition.

Relevance is an expression of the fitting of a certain search result to a domain of interest chosen a priori.

Unexpectedness is an expression of the rarity or surprising properties of a certain search result relative to the common properties of the referred domain of interest.

In this paper we deal with two Interestingness measures obeying the definition in formula (1).

### 2.1  Match-Mismatch

Our first Interestingness measure, proposed in the same ref. by Exman [1], is the match-mismatch function:

$$Int = Match * Mismatch \qquad (2)$$

where * is just the numerical multiplication operator.

Match is a numerical expression of the fitting between a specific search result and a formal characterization of the chosen domain of interest. This characterization can be given by a set of keywords or by an ontology of the domain. The numerical expression can be say, Boolean or Integer.

Mismatch is a numerical expression of the lack of fitting of a specific search result and a formal characterization of the referred domain. Mismatch points out to results that, while still belonging to the chosen domain of interest, they are not the most typical ones. For instance, mercury does conduct heat and electricity as all metals, but it is *not* a typical metal, since it is liquid, in contrast to most metals that are solid at room temperature.

### 2.2  Tf-Idf

Another measure of interestingness obeying definition (1) is the well-known Tf-Idf function, which has been used for information retrieval:

$$Int = Tf * Idf \qquad (3)$$

*Tf* is the term frequency, viz. the frequency of appearance of a term in a given search result. The higher the frequency of a term that belongs to the characterization of a domain of interest, the more relevant is the result to the chosen domain.

*Idf* is the inverse document frequency. In our case a document is a search result containing terms belonging to the domain of interest. The rarest are such documents, the most interesting they are.

## 3  The Interestingness Tool – Software Architecture

In this section we overview the software architecture of a flexible and generic tool for web search based on Interestingness functions.

We first deal with architectural considerations. The Interestingness Tool should be flexible and generic along a series of dimensions.

The tool is supposed to work based upon rearrangement of the order of results obtained from commercial search engines. Therefore, the tool should be independent of particular search engines and accept search engines as parameters.

In order to calculate the Interestingness function values for search results, the tool needs files containing characterizations of a chosen domain of interest. The tool should be able to read files provided by external sources.

The internal mechanisms of the tool should be capable to work with different Interestingness functions. One should easily plug-in new functions, without having to re-write any other parts of the referred mechanisms.

Finally, the tool should be accessible as a client from anywhere in the internet.

### 3.1    Overall Architecture

Following the above considerations the Interestingness Tool is actually built in a client-server architecture to allow its use throughout the internet.

The tool client displays a Graphical User Interface (GUI).

The respective Server has two well separated upper level modules:

- Web Search – with pluggable API's for different Search Engine API's;
- Interestingness functions – with independently pluggable algorithms.

The interestingness functions' module performs reordering of search results obtained by the Web Search module.

A schematic overall architecture of the Interestingness tool is shown in Figure 1.

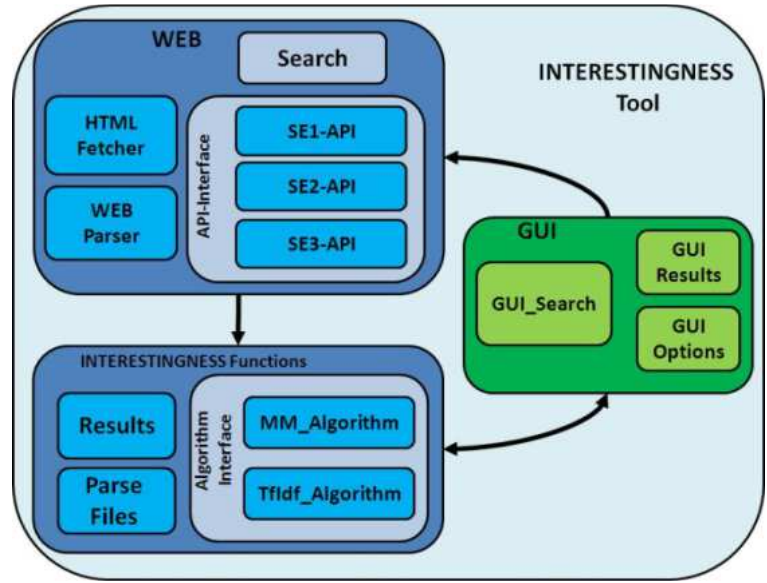

**Fig. 1.** The Interestingness Tool Software Architecture – There are three upper level modules: a client GUI (green) at the r.h.s.; WEB search and Interestingness Functions in the Server (blue) at the l.h.s. The WEB search module has an API-interface for each of the different Search Engines (SE). The Interestingness Functions have an interface for Match-Mismatch (MM) and TfIdf algorithms.

## 4  The Interestingness Tool - Implementation

In this section we describe implementation decisions for the Interestingness Tool.

The tool was developed in a client-server format, and encoded in the C# language. There are two input files for each chosen domain of interest. These are:

- *Target file* – contains the characterization in terms of keywords, possibly from a given ontology, of the domain of interest;
- *Competitor file* – contains terms that may belong to the domain of interest, but may also appear in other domains; the purpose of this file is to avoid domain ambiguity.

A third file needed – which is not specific to a domain of interest – is a *Stopwords* file. These are generic words that do not contribute any specificity to the Interestingness calculation and are eliminated.

The Interestingness Tool                                Iaakov ExmanThe target and competitor files are especially prepared for each domain. They may evolve in time as a consequence of experience gained with the tool. The stopwords file used is downloaded from the internet.

### 4.1 Interestingness Function Computations

Given the above files, the Match-Mismatch computation is an extension of equation (2) as follows:

$$Int = \frac{(Match * Mismatch) - Competitors}{Normf} \quad (4)$$

where *Competitors* is a sum of the appearances in the given search result, of keywords found in the Competitor file; *Normf* is a normalization factor needed to make the computation independent of the size of the given search result.

The *Match* value is a sum of the appearances in the given search result, of keywords found in the Target file.

The *Mismatch* value is computed by the symmetric difference expression between the set of terms in the given search result $S$ and the set of terms in the Target file $T$:

$$Mismatch = T \Delta S = (T - S) \cup (S - T) \quad (5)$$

For the TfIdf function, one takes into account the Target file for the computation of *Tf*, in a very similar fashion to the computation of the Match value.

The computation of $Idf_k$ for search results containing a given keyword $k$ is given as usual by:

$$Idf_k = \log(N/df_k) \quad (6)$$

where $N$ is the total number of search results under consideration and $df_k$ is the number of search results containing the keyword k.

## 5 Validation

Here we describe a sample of experimental results. These are obtained as part of preliminary investigations with the Interestingness Tool, beyond normal quality assurance of the tool. Essentially, below are shown two kinds of comparisons for the order of search results:

- The Interestingness Tool vs. commercial search engines – such as Google, Bing and Yahoo;
- Different Interestingness Functions – say, Match-Mismatch vs. TfIdf.

These comparisons were done for a variety of domains of interest. Here only a small sample of the domains is displayed.

### 5.1    Experimental Results

We first deal with comparisons of the Interestingness Tool with commercial search engines. Then we compare Interestingness functions. Fig. 2 displays a partial Target File for the chosen domain of interest *Space Stars* in astronomy.

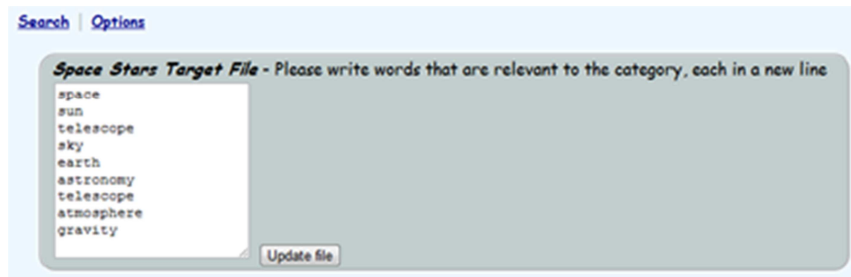

**Fig. 2.** Target File input for Space Stars – This is a partial Target File manually input into the GUI of the Interestingness Tool. One can also automatically input whole files.

**The Interestingness Tool vs. Commercial Search Engines.**

Table 1 shows the first ten search results given by the Interestingness Tool (with the TfIdf function) against the order of the same results given by Google search. The interest domain is "Space Stars" with search keyword "Mars". The same results are graphed in Fig. 3.

Table 1: Search Results Order – Space Stars.

| Interestingness Tool | Google |
|---|---|
| 1 | 133 |
| 2 | 310 |
| 3 | 99 |
| 4 | 40 |
| 5 | 614 |
| 6 | 498 |
| 7 | 18 |
| 8 | 44 |
| 9 | 334 |
| 10 | 181 |

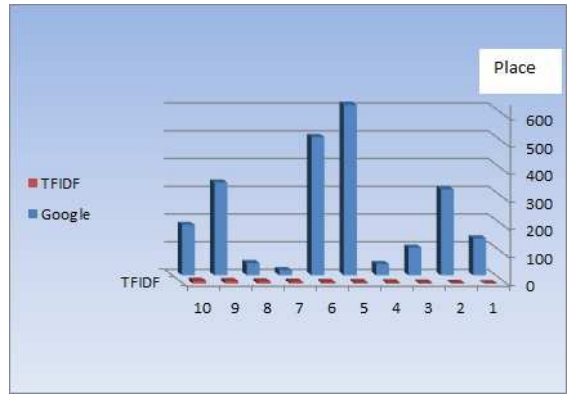

**Fig. 3.** Search Results Order for "Space Stars" – In the small (red) front columns are the Interestingness results. In the big (blue) back columns one sees the respective Google results.

One can see that the reordering of the Interestingness Tool is very significant. For instance, result 5 in the Interestingness Tool, which is calculated as very interesting, is found in the place 614 in the Google search. This implies that the chances of a person to manually find this result by quickly reading the search result pages are minimal.

**Comparison of Tf-Idf with Match-Mismatch.**

We now show results for the comparison of the Match-Mismatch function with the TfIdf function, for different domains of interest.

Table 2 shows the order of the first ten search results given by the Interestingness Tool, comparing the order of Match-Mismatch function against the order of the same results given by the TfIdf function. The domain of interest is "Musicals" and the search keyword was "Cats".

Table 2: Search Results Order – Musicals.

| Match-Mismatch | TfIdf |
|---|---|
| 1 | 2 |
| 2 | 1 |
| 3 | 3 |
| 4 | 5 |
| 5 | 4 |
| 6 | 6 |
| 7 | 10 |
| 8 | 9 |
| 9 | 8 |
| 10 | 7 |

The same results are graphically shown in Fig. 4. One sees that the order of TfIdf results is just a simple permutation of the order of the Match-Mismatch results, with a similar overall trend.

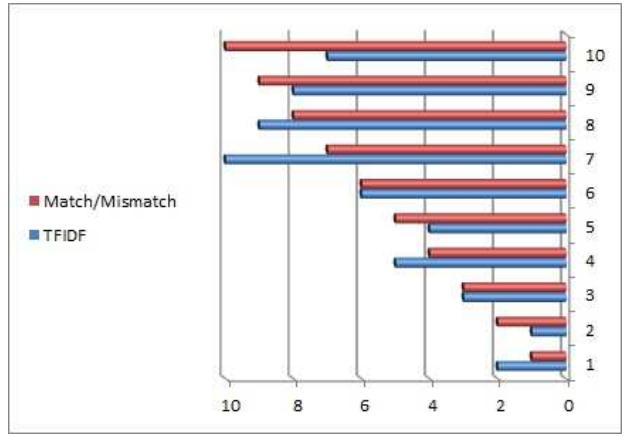

**Fig. 4.** Search Results Order for "Musicals" – The upper (red) horizontal in each pair of columns shows the Match-Mismatch results. In the lower (blue) horizontal columns one sees the respective TfIdf results.

Table 3 again shows the order of the first ten search results given by the Interestingness Tool, comparing the order of Match-Mismatch function against the order of the same results given by the TfIdf function. Now the domain of interest is "Space Stars" and search keyword was "Mars".

Table 3: Search Results Order – Space Stars.

| Match-Mismatch | TfIdf |
|---|---|
| 1 | 1 |
| 2 | 3 |
| 3 | 5 |
| 4 | 2 |
| 5 | 7 |
| 6 | 4 |
| 7 | 20 |
| 8 | 6 |
| 9 | 14 |
| 10 | 15 |

The same results are graphically shown in Fig. 5. The order of TfIdf results is almost a permutation of the order of the Match-Mismatch results, up to the 6$^{th}$ place. From the 7$^{th}$ until the 10$^{th}$ place there are slightly higher positions, but still with a difference of the order of 10 places.

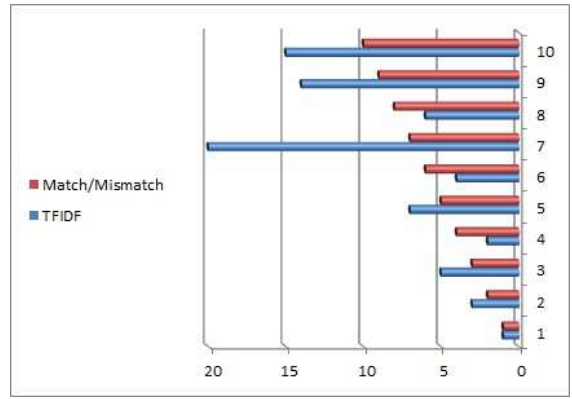

**Fig. 5**. Search Results Order for "Space Stars" – The upper (red) horizontal in each pair of columns shows the Match-Mismatch results. In the lower (blue) horizontal columns one sees the respective TfIdf results.

Finally, Table 4 again shows the order of the first ten search results, comparing the order of Match-Mismatch function against the order of the same results given by the TfIdf function. Now the domain of interest is "Basketball" and search keyword was "Michael", just a frequent name for basketball players.

Table 4: Search Results Order – Basketball.

| Match-Mismatch | TfIdf |
|---|---|
| 1 | 1 |
| 2 | 2 |
| 3 | 3 |
| 4 | 5 |
| 5 | 4 |
| 6 | 6 |
| 7 | 8 |
| 8 | 7 |
| 9 | 9 |
| 10 | 560 |

The same results are graphically shown in Fig. 6, discarding the last row in the table which is an obvious outlier.

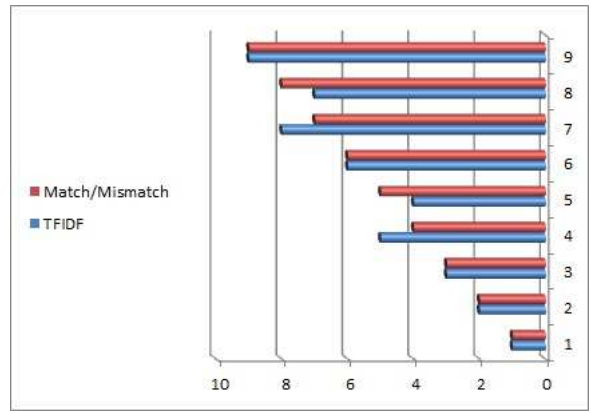

**Fig. 6**. Search Results Order for "Basketball" – The upper (red) horizontal in each pair of columns shows the Match-Mismatch results. In the lower (blue) horizontal columns one sees the respective TfIdf results.

One sees that the order of TfIdf results is once again an exact permutation of the order of the Match-Mismatch results, up to the $9^{th}$ place. The $10^{th}$ result is an outlier, as stated above.

Thus, results from three completely different domains consistently show that both Match-Mismatch and TfIdf can serve as interchangeable Interestingness functions.

## 6   Discussion

This paper has described a tool for Interestingness based search of the web.

The software architecture of the Interestingness Tool is such that one obtains flexibility and robustness with respect to several dimensions.

The tool was validated by a series of experiments for a variety of domains of interest, diverse interestingness functions and against most common commercial search engines.

An important result obtained by this preliminary investigation is that the generic definition of Interestingness, as a composition of Relevance and Unexpectedness functions, is corroborated by the consistent equivalence of quite different functions as Match-Mismatch and TfIdf, as seen for diverse domains of interest.

### 6.1   Future Work

A fundamental issue to be investigated is the demonstration of the correctness of the empirical results obtained, by means of an Interestingness criterion independent of the idea of composition of Relevance and Unexpectedness.

The Interestingness Tool                                  Iaakov Exman

It is worthwhile to propose new Interestingness functions. It would be desirable to have some systematic approach to generate such functions.

We plan to significantly extend the current preliminary investigation, by opening the Interestingness Tool for general use as a Web Service, and accumulate evidence by collaboration with users outside of our group. In particular, we intend to make comparisons with additional search engines.

### 6.2  Main Contribution

The main contribution of this work is an empirically demonstrated Interestingness Tool for web search. It displays enough flexibility and robustness to be a production tool for thorough and systematic investigation of Interestingness issues.

## References

1. Exman, I., "Interestingness – A Unifying Paradigm – Bipolar Function Composition", in Fred, A. (ed.) Proc. KDIR'2009 Int. Conference on Knowledge Discovery and Information Retrieval, pp. 196-201, (2009). DOI = 10.5220/0002308401960201, e-print arXiv:1404.0091 [cs.IR], (April 2014).

2. Exman, I., "Web Search of New Linearized Medical Drug Leads", original SKY'2011 2$^{nd}$ Int. Workshop on Software Knowledge, pp. 108-115, October 2011, SciTePress, Portugal, DOI = 10.5220/0003705401080115, e-print arXiv:1404.3435 [cs.IR], (April 2014).

3. Geng, L. and Hamilton, H.J., "Interestingness measures for data mining: A survey", ACM Computing Surveys, Vol. 38, (3), article N0. 9 (2006). DOI = 10.1145/1132960.1132963.

4. Geng, L. and Hamilton, H.J., "Choosing the Right Lens: Finding What is Interesting in Data Mining", in Ref. [Guillet and Hamilton, 2007], pp. 3-24, (2007).

5. Guillet, F. and Hamilton, H.J., *Quality Measures in Data Mining,* Sutdies in Computational Intelligence, Vol. 43, Springer, Heidelberg, Germany, (2007).

6. Huynh, X-H., Guillet, F., Blanchard, J., Kuntz, P., Briand, H. and Gras R., "A Graph-based Clustering Approach to Evaluate Interestingness Measures: A Tool and a Comparative Study", pp. 25-50, in (Guillet and Hamilton, ref. [4]) (2007). DOI = 10.1007/978-3-540-44918-8_2.

7. Huynh, X-H., Guillet, F. and Briand, H., 'ARQAT: An Exploratory Analysis Tool For Interestingness Measures", (2005).

8. Klosgen, W. and Zytkow, J.M. (eds.), *Handbook of Data Mining and Knowledge Discovery*, Oxford University Press, Oxford, Japan, (2002).

9. Lin, T., Etzioni, O. and Fogarty, J., "Identifying Interesting Assertions from the Web", in Proc. CIKM'09, 18[th] ACM Conf. on Information and Knowledge Management, pp. 1787-1790, (2009). DOI = 10.1145/1645953.1646230.

10. McGarry, K., "A survey of interestingness measures for knowledge discovery", *Knowledge Engineering Review J.*, 20 (1), pp. 39-61, (2005). DOI = 10.1017/S0269888905000408.

11. Padmanabhan, B. and Tuzhilin, A., "Unexpectedness as a measure of interestingness in knowledge discovery", *Decision Support Sys.*, Vol. 27, (3), pp. 303-318, (December 1999). DOI = 10.1016/S0167-9236(99)00053-6.

12. Piatetsky-Shapiro, G. and Matheus, C.J., "The Interestingness of Deviations", KDD-94, AAAI-94 Knowledge Discovery in Databases Workshop, (1994).

13. Tuzhilin, A., "Usefulness, Novelty, and Integration of Interestingness Measures", chapter 19.2.2 in reference [Klosgen and Zytkow, 2002], pp. 496-508, (2002).

14. Webb, G.I., "Preliminary Investigations into Statistically Valid Exploratory Rule Discovery", in AUSDM03, Proc. 2003 Australasian Data Mining Workshop, pp. 1-9, (2003).